# A sectorial approach to Kaluza-Klein theory

Terence V. Sewards
Sandia Research Center
21 Perdiz Canyon Road
Placitas, N.M.  87043

(505) 867-4392
tsewards@cognisurf.com

**Abstract**

   The Kaluza and Klein versions of Kaluza-Klein theory are reviewed and compared. The differences in the field equations of the two theories are related to the transformation properties of the metrics employed. Based on this comparison a modified version of Kaluza's theory is proposed, in which the different sectors of the metric which describe gravitation, electromagnetism and the scalar field are considered to be partially separate entities, with distinct coordinate transformation properties. In addition, the "cylinder condition" is relaxed, so that it only affects the 4-D sector of the metric. This results in a theory in which the gravitational and electromagnetic fields are restricted to a four-dimensional "brane", and the scalar field is restricted to the 5-D "bulk".

## I. INTRODUCTION

   Kaluza-Klein (KK) theory seeks to unify two of the fundamental forces of nature, gravitation and electromagnetism. In Kaluza's original paper[1], the 4-D theory of general relativity (GR) derived originally by Einstein[2] is extended to a 5-D spacetime. The resulting equations are then separated into three sets, one of which is equivalent to the Einstein 4-D field equations for gravitation, another to Maxwell's equations for the electromagnetic (EM) field, and a third one which describes a scalar field. Independently, Klein[3] elaborated a 5-D extension of GR, but with significant differences. Here these two theories are compared and contrasted, and a variant of Kaluza's original formulation, in which the cylinder condition is partially relaxed, is elaborated. Reviews of the theoretical bases of Kaluza-Klein theory, in which both the Kaluza and Klein formulations are examined, include those of Chyba[4], Applequist et al.[5], Bailin and Love[6], Pasini[7], Coquereaux and Esposito-Farèse[8], Schwarz and Doughty[9], Duff[10], Overduin and Wesson[11], O'Raifeartaigh and Straumann[12], van Dongen[13], Goenner[14], and Ivanov and Prodanov[15]. Reviews explicitly devoted to Kaluza's theory include those of Fabbri[16], Atondo-Rubio et al.[17], and Kumar and Suresh[18]. Here and below, a "brane/bulk" approach to Kaluza-Klein theory is taken, wherein the 4-D sector of the KK metric is considered to reside on a 4-D brane, while the 5-D sector resides in the bulk.

## II. KALUZA'S 5-D FORMULATION

   In Kaluza's theory the five dimensional line element is written as

$$d\hat{s}^2 = \hat{g}_{\hat{\mu}\hat{\nu}}(x^\mu, x^5) d\hat{x}^{\hat{\mu}} d\hat{x}^{\hat{\nu}} \qquad (1)$$

where $\hat{\mu}, \hat{\nu} = 1, 2, 3, 4, 5$, and $x^5$ is the additional spatial coordinate. The 5-D metric corresponding to this line element is given by

$$\hat{g}_{\hat{\mu}\hat{\nu}} = \begin{bmatrix} g_{\mu\nu} & g_{\mu 5} \\ g_{5\nu} & g_{55} \end{bmatrix} \qquad (2)$$

Here $g_{\mu\nu}$ is the ordinary 4-D GR metric tensor, $\mu = 1, 2, 3, 4$, and the fourth coordinate is the temporal one. To account for the fact that physical qualities depend on the usual 4-D spacetime manifold, Kaluza required that 5-D metric tensor satisfy the "cylinder condition":

$$g_{\hat{\mu}\hat{\nu},5} = 0 \qquad (3)$$

(i.e., derivatives of all the components of the metric with respect to the fifth dimension vanish). The 5-D Christoffel symbols of the first kind are assumed to have the same form as their 4-D counterparts, and are given by

$$2\Gamma_{\hat{\lambda}\hat{\mu}\hat{\nu}} = g_{\hat{\lambda}\hat{\nu},\hat{\mu}} + g_{\hat{\lambda}\hat{\mu},\hat{\nu}} - g_{\hat{\mu}\hat{\nu},\hat{\lambda}} \qquad (4)$$

Symbols involving the fifth coordinate are calculated employing the cylinder condition, expressed in equation (3):

$$2\Gamma_{\lambda\mu 5} = 2\Gamma_{\lambda 5\mu} = g_{\lambda 5,\mu} - g_{\mu 5,\lambda} \qquad (5)$$

$$2\Gamma_{5\mu\nu} = g_{5\nu,\mu} + g_{5\mu,\nu} \qquad (6)$$

$$2\Gamma_{55\mu} = g_{55,\mu} \qquad (7)$$

$$2\Gamma_{5\mu 5} = g_{55,\mu} \qquad (8)$$

$$2\Gamma_{\mu 55} = 2g_{\mu 5,5} - g_{55,\mu} \qquad (9)$$

$$2\Gamma_{555} = 0 \qquad (10)$$

At this point, the components $g_{5\mu}$ and $g_{55}$ are set as

$$g_{5\mu} = 2\alpha A_\mu \qquad (11)$$

$$g_{55} = 2\psi \qquad (12)$$

where $\alpha$ is a multiplicative constant, and $\psi$ is a scalar field. In Kaluza's formulation, the terms $A_\mu$ are identified as the vector and scalar electromagnetic potentials $[A_x, A_y, A_z, \phi_e/c]$. Equations (5), (6) and (7) now become

$$\Gamma_{\lambda\mu 5} = \Gamma_{\lambda 5\mu} = \alpha(A_{\lambda,\mu} - A_{\mu,\lambda}) = \alpha F_{\lambda\mu} \tag{13}$$

$$\Gamma_{5\mu\nu} = \alpha(A_{\mu,\nu} + A_{\nu,\mu}) = \alpha \Sigma_{\mu\nu} \tag{14}$$

$$\Gamma_{55\mu} = \Gamma_{5\mu 5} = \psi_{,\mu} \tag{15}$$

The $F_{\nu\sigma}$ terms are related through the equation

$$F_{\mu\nu,\lambda} + F_{\nu\lambda,\mu} + F_{\lambda\mu,\nu} = 0 \tag{16}$$

It is apparent from the above relationships that the terms $F_{\mu\nu}$ can be identified as the components of the covariant electromagnetic field tensor.

The Christoffel symbols of the second kind, which are necessary in order to construct the Riemann-Christoffel and Ricci tensors, cannot be obtained directly from those of the first kind, since by definition

$$\Gamma^{\alpha}{}_{\mu\nu} = g^{\alpha\lambda} \Gamma_{\lambda\mu\nu} \tag{17}$$

and the elements $g^{\alpha\lambda}$ can only be determined by inversion of the matrix $[g_{\alpha\lambda}]$, which is unknown. Thus, as in 4-D GR theory, a general solution to Kaluza's 5-D formulation cannot be obtained, and one must either employ the linearized (weak-field) approximation, or attempt to find an exact solution by specifying a particular geometry, as is employed in obtaining the Schwarzschild solution in GR. Kaluza chose the former option, whereby the 5-D metric $g_{\hat{\mu}\hat{\nu}}$ is expressed in terms of linear deviations $h_{\hat{\mu}\hat{\nu}}$ from the 5-D Minkowski metric $\eta_{\hat{\mu}\hat{\nu}}$ as

$$g_{\hat{\mu}\hat{\nu}} \approx \eta_{\hat{\mu}\hat{\nu}} + h_{\hat{\mu}\hat{\nu}} \qquad \text{with} \qquad |h_{\hat{\mu}\hat{\nu}}| \ll 1 \tag{18}$$

The Christoffel symbols of the second kind may now be obtained from those of the first kind using the relations

$$\Gamma^{\hat{\alpha}}{}_{\hat{\mu}\hat{\nu}} \approx \eta^{\hat{\alpha}\hat{\lambda}} \Gamma_{\hat{\mu}\hat{\nu}\hat{\lambda}} = -\Gamma_{\hat{\mu}\hat{\nu}\hat{\alpha}} \tag{19}$$

The weak-field 5-D Riemann-Christoffel curvature tensor is given by the relations

$$R^{\hat{\mu}}{}_{\hat{\nu}\hat{\lambda}\hat{\sigma}} = \Gamma^{\hat{\mu}}{}_{\hat{\nu}\hat{\sigma},\hat{\lambda}} - \Gamma^{\hat{\mu}}{}_{\hat{\nu}\hat{\lambda},\hat{\sigma}} \tag{20}$$

in which terms involving products of Christoffel symbols have been neglected. The corresponding 5-D Ricci array is

$$R_{\hat{\mu}\hat{\nu}} = \Gamma^{\hat{\alpha}}{}_{\hat{\mu}\hat{\nu},\hat{\alpha}} - \Gamma^{\hat{\beta}}{}_{\hat{\mu}\hat{\beta},\hat{\nu}} \tag{21}$$

The 4-D (gravitational) Ricci tensor is then

$$R_{\mu\nu} = \Gamma^{\alpha}{}_{\mu\nu,\alpha} - \Gamma^{\beta}{}_{\mu\beta,\nu} \qquad (22)$$

The field equations that describe Kaluza's 5-D unified gravity and electromagnetism are then:

*Gravity:* $\qquad R_{\mu\nu} = -\kappa(T_{\mu\nu} - \frac{1}{2}\eta_{\mu\nu}T) \qquad (23)$

*EM:* $\qquad R_{5\mu} = -\kappa T_{5\mu} \qquad (24)$

*Scalar field:* $\qquad R_{55} = -\Box\psi = -\kappa\left(T_{55} - \frac{1}{2}\eta_{55}T\right) \qquad (25)$

Here $\kappa = 8\pi G/c^2$ is the gravitational constant, and $T_{\mu\nu}$ is the 4-D energy-momentum array for the mass source. The $R_{5\mu}$ (electromagnetic) components can be expressed in terms of the electromagnetic potentials or the covariant electromagnetic field tensor:

$$R_{5\nu} = -\alpha(A_{\sigma,\nu} - A_{\nu,\sigma})_{,\sigma} = -\alpha F_{\sigma\nu,\sigma} \qquad (26)$$

and the gravitational equation (23) may be written in terms of the Einstein tensor

$$G_{\mu\nu} = R_{\mu\nu} - \frac{1}{2}\eta_{\mu\nu}R = -\kappa T_{\mu\nu} \qquad (27)$$

The components of the 4-D Ricci tensor are now expressed in terms of the metric deviations in the form

$$R_{\mu\nu} = \frac{1}{2}\left(h_{\mu}{}^{\beta}{}_{,\nu\beta} + h_{\nu}{}^{\beta}{}_{,\mu\beta} - h_{\mu\nu,\beta}{}^{\beta} - h_{,\mu\nu}\right) \qquad (28)$$

with $h$ being the trace of the metric deviation tensor $h_{\mu\nu}$[19]. Defining the "trace-reversed" metric deviation tensor

$$\bar{h}_{\mu\nu} = h_{\mu\nu} - \frac{1}{2}\eta_{\mu\nu}h \qquad (29)$$

the Einstein tensor becomes

$$G_{\mu\nu} = -\frac{1}{2}\left(\bar{h}_{\mu\nu,\beta}{}^{\beta} + \eta_{\mu\nu}\bar{h}_{\beta\gamma,}{}^{\beta\gamma} - \bar{h}_{\mu\beta,}{}^{\beta}{}_{\nu} - \bar{h}_{\nu\beta,}{}^{\beta}{}_{\mu}\right) \qquad (30)$$

Both $h_{\mu\nu}$ and $\bar{h}_{\mu\nu}$ transform as components of a tensor in 4-D flat spacetime[19]. After imposing the harmonic gauge $\bar{h}^{\mu\nu}_{,\nu}=0$ one obtains the 4-D field equations of linearized gravitation in terms of the trace-reversed deviations:

$$\Box_{(4)}\bar{h}_{\mu\nu} = -2G_{\mu\nu} = 2\kappa T_{\mu\nu} \tag{31}$$

in which $\Box_{(4)}$ is the 4-D d'Alembertian operator, given by

$$\Box_{(4)} = \frac{1}{c^2}\frac{\partial^2}{\partial t^2} - \nabla^2 \tag{32}$$

Similarly, the electromagnetic terms can be expressed in terms of the deviations (field potentials) $\bar{h}_{\mu 5}$:

$$G_{\mu 5} = R_{\mu 5} = -\frac{1}{2}\Box_{(4)}\bar{h}_{\mu 5} = -\kappa T_{\mu 5} \tag{33}$$

Employing equations (11), (18) and (29) one obtains

$$\bar{h}_{\mu 5} = 2\alpha A_\mu \tag{34}$$

and

$$\bar{h}_{55} = 2\psi \tag{35}$$

The potentials for all three fields can now be written in 5-D matrix form as

$$\bar{h}_{\hat{\mu}\hat{\nu}} = \begin{bmatrix} \bar{h}_{11} & \bar{h}_{12} & \bar{h}_{13} & \bar{h}_{14} & 2\alpha A_x \\ \bar{h}_{12} & \bar{h}_{22} & \bar{h}_{23} & \bar{h}_{24} & 2\alpha A_y \\ \bar{h}_{13} & \bar{h}_{23} & \bar{h}_{33} & \bar{h}_{34} & 2\alpha A_z \\ \bar{h}_{14} & \bar{h}_{24} & \bar{h}_{34} & \bar{h}_{44} & 2\alpha\phi_e/c \\ 2\alpha A_x & 2\alpha A_y & 2\alpha A_z & 2\alpha\phi_e/c & 2\psi \end{bmatrix} \tag{36}$$

The 4-D gravitational field components are contained in the first four rows and columns of the array (36), the electromagnetic potentials in the fifth row and column, and the scalar field occupies the (5,5) position in the array. This array is not a tensor, and Kaluza did not discuss its transformation properties in his 5-D formulation. Einstein and Bergmann[20] examined these properties in some depth, but obtained a formulation that is closer to Klein's than Kaluza's.

At this point a general expression for all three field potential equations has been obtained in terms of the trace-reversed deviations:

$$\Box_{(4)} \bar{h}_{\hat{\mu}\hat{\nu}} = 2\kappa T_{\hat{\mu}\hat{\nu}} \qquad (37)$$

The EM and scalar field compartments are thus

$$\Box_{(4)} \bar{h}_{\mu 5} = 2\kappa T_{\mu 5} \qquad (38)$$

$$\Box_{(4)} \bar{h}_{55} = 2\kappa T_{55} \qquad (39)$$

If the gravitational source consists of a distribution of non-interacting moving matter ("dust"), the components of the 4-D source (energy-momentum) tensor are given by $T_{\mu\nu} = \rho_m u_\mu u_\nu$. Kaluza extends this to 5-D through the specification $T_{\hat{\mu}\hat{\nu}} = \rho_m u_{\hat{\mu}} u_{\hat{\nu}}$. The source array then has the form

$$T_{\hat{\mu}\hat{\nu}} = \begin{bmatrix} \rho_m u_x^2 & \rho_m u_x u_y & \rho_m u_x u_z & \rho_m c u_x & \rho_m c_5 u_x \\ \rho_m u_y u_x & \rho_m u_y^2 & \rho_m u_y u_z & \rho_m c u_y & \rho_m c_5 u_y \\ \rho_m u_z u_x & \rho_m u_z u_y & \rho_m u_z^2 & \rho_m c u_z & \rho_m c_5 u_z \\ \rho_m c u_x & \rho_m c u_y & \rho_m c u_z & \rho_m c^2 & \rho_m c c_5 \\ \rho_m c_5 u_x & \rho_m c_5 u_y & \rho_m c_5 u_z & \rho_m c c_5 & \rho_m c_5^2 \end{bmatrix} \qquad (40)$$

As in the case of the field array, this object is not a tensor. If the velocity of the source matter is low (i.e. $|\mathbf{u}| \ll c$), terms of the form $\rho_m u_i u_j$ may be neglected. In addition, Kaluza specifies that $c_5 \ll c$, so that the trace of $T_{\mu\nu}$ is $T = -\rho_m c^2$. However, this approximation is untenable (see below). It may be seen that in the array (40) the sources of the electromagnetic field $T_{5\mu} = (\rho_m c_5 u_i, \rho_m c c_5)$ do not contain the charge density or current overtly, but rather depend on the mass density $\rho_m$. Kaluza remedies this problem by arguing that the velocity $c_5$ contains the source particle charge.

Equations (38) must correspond to the field equations for the electromagnetic scalar and vector potentials, expressed here in Gaussian units:

$$\frac{1}{c^2} \frac{\partial^2 \phi_e}{\partial t^2} - \nabla^2 \phi_e = 4\pi \rho_e \qquad (41)$$

$$\frac{1}{c^2} \frac{\partial^2 A_i}{\partial t^2} - \nabla^2 A_i = \frac{4\pi}{c} J_i^e \qquad (42)$$

This correspondence requires

$$\rho_e = 2\alpha c c_5 \rho_m \qquad (43)$$

which establishes the relationship between the velocity $c_5$ and the charge (density). Kaluza mentions that Einstein pointed out to him that the velocity $c_5$ for electrons calculated from this relationship is "enormously large" (grossly superluminal). However, $c_5$ is not the source matter (electron, proton) velocity, which is given by $(u_x, u_y, u_z)$, but rather the fifth component of the proper velocity, the fourth component being the velocity of light $c$. The latter velocity is a property of the 4-D vacuum, and characterizes the propagation of massless bosons in ordinary 4-D space. By analogy, it is reasonable to suppose that $c_5$ is a property of the 5-D bulk, and characterizes the propagation of massless bosons in that space. As Gonzalez-Mestres[21] points out, if one takes a sectorial approach to Lorentz covariance, this covariance only affects particles within a given sector of spacetime (e.g. a 4-D "brane"), and it is unreasonable to expect that the characteristic velocities in other sectors (e.g. the 5-D "bulk") should be the same as in the 4-D sector.

Together, equations (36), (37) and (40) contain Einstein's gravitational (GR) equations and Maxwell's equations, thus "unifying" gravity and electromagnetism into a single set of equations. It should be noted, however, that this unification does not in itself imply that gravity and electromagnetism are in some manner part of a larger or global physical system, as the equations for each field are entirely independent.

The equation for the scalar field is:

$$\Box_{(4)} \psi = \frac{\kappa c^2}{2} \rho_m \tag{44}$$

in which $\rho_m$ is again the mass density. It should be noted that this equation is based on the approximation $c_5 \ll c$, which - as mentioned above - is not tenable.

The Kaluza metric is a 5 x 5 symmetric array, so it has 15 independent components. Hence there are 5 degrees of freedom, which correspond to five particles (massless bosons)[4,18]. These include two spin-2 gravitons, two spin-1 photons, and a spin-0 scalar boson.

In order to derive the equation of motion, Kaluza's starts with a 5-D version of the 4-D geodesic equation

$$\frac{du^\sigma}{ds} + \Gamma^\sigma{}_{\hat\mu\hat\nu} u^{\hat\mu} u^{\hat\nu} = 0 \tag{45}$$

which is expanded into its component parts:

$$\frac{du^\sigma}{ds} + \Gamma^\sigma{}_{\mu\nu} u^\mu u^\nu + 2\Gamma^\sigma{}_{\mu 5} c\, c_5\, u^\mu + \Gamma^\sigma{}_{55} c_5^2 = 0 \tag{46}$$

The first two terms are just the geodesic equation of 4-D relativistic gravity, the third term should correspond to the (Lorentz) electromagnetic force, and the fourth is due to the scalar field. Substitution of the values for the two 5-D Christoffel symbols in (46) yields

$$\frac{du^\sigma}{ds} + \Gamma^\sigma{}_{\mu\nu} u^\mu u^\nu + 2\alpha F_{\mu\sigma} c c_5 u^\mu + \psi_{,\sigma} c_5^2 = 0 \tag{47}$$

The fourth term in equation (47) is problematic because of the large value of $c_5$, as pointed out by Einstein to Kaluza[1]. The Lorentz force for a test particle with inertial mass $m_{ti}$ and charge $q_{te}$ is given by

$$m_{ti} \frac{du^\sigma}{ds} = q_{te} F_{\sigma\mu} u^\mu \tag{48}$$

Comparison of (47) and (48) validates equation (43) above.

### III. KLEIN'S 5-D FORMULATION

In his version of the 5-D theory, Klein[3] starts with the 5-D line element

$$d\sigma^2 = \gamma_{\hat{\mu}\hat{\nu}} dx^{\hat{\mu}} dx^{\hat{\nu}} \tag{49}$$

where again $\hat{\mu}, \hat{\nu} = 1, 2, 3, 4, 5$, and $x^5$ is the additional spatial coordinate. Like Kaluza, Klein invokes the cylinder condition, by which the quantities $\gamma_{\hat{\mu}\hat{\nu}}$ are not to depend on the fifth coordinate $x^5$. However, unlike Kaluza, Klein does not take the 4-D elements $\gamma_{\mu\nu}$ to be equal to the $g_{\mu\nu}$ of GR. Rather, he requires that the coordinates $x^\mu$ characterize the usual 4-D spacetime. This means that the coordinates must transform between 4-D frames according to the usual transformation law

$$x^\mu = f^\mu(x^{\nu'}) \tag{50}$$

Permitted coordinate transformations are restricted to

$$x^\mu = \varphi_\mu(x^{1'}, x^{2'}, x^{3'}, x^{4'}) \tag{51}$$

$$x^5 = x^{5'} + \varphi_0(x^{1'}, x^{2'}, x^{3'}, x^{4'}) \tag{52}$$

The metric element $\gamma_{55}$ is evidently invariant under this transformation, and Klein sets it to a constant. In 4-D spacetime, the line element $ds^2 = g_{\mu\nu} dx^\mu dx^\nu$ between two events $x$ and $x + dx$ may be split into separate components, $ds^2 = g_{44} d\lambda^2 + dl^2$, where the temporal and spatial differentials are given by

$$d\lambda = dx^4 + \frac{g_{i4}}{g_{44}} dx^i \tag{53}$$

$$dl^2 = \left(g_{ij} - \frac{g_{i4} g_{j4}}{g_{44}}\right) dx^i dx^j \tag{54}$$

In 5-D spacetime an analogous decomposition, in which the cylinder condition is incorporated, may be effected [4,9]. Thus two events are separated by the interval $d\sigma^2 = \gamma_{55} d\theta^2 + ds^2$, where the quantities

$$d\theta = dx^5 + \left(\frac{\gamma_{5\mu}}{\gamma_{55}}\right) dx^\mu \tag{55}$$

$$ds^2 = \left(\gamma_{\mu\nu} - \frac{\gamma_{5\mu} \gamma_{5\nu}}{\gamma_{55}}\right) dx^\mu dx^\nu \tag{56}$$

are invariants, so that $d\sigma^2$ is also invariant.

In order to illustrate the more general formulation of Klein's theory, rather than setting $\gamma_{55}$ to be a constant, the choice $\gamma_{55} = \phi_K^2$ is made, with the scalar field $\phi_K$ a function of $x$, $y$, $z$ and $t$, but not of $x_5$. The 5-D metric components are set as $\gamma_{5\mu} = \kappa \phi_K A_\mu$, so the 5-D metric tensor becomes

$$\gamma_{\hat\mu\hat\nu} = \begin{bmatrix} g_{\mu\nu} + \kappa^2 \phi_K^2 A_\mu A_\nu & \kappa^2 \phi_K^2 A_\mu \\ \kappa^2 \phi_K^2 A_\nu & \phi_K^2 \end{bmatrix} \tag{57}$$

This metric is expressed exclusively in terms of the 4-D quantities $g_{\mu\nu}$ and $A_\mu$, and the scalar potential $\phi_K$. Field equations are then obtained by variation of the 5-D action

$$S = \int d^5 x \sqrt{-\gamma} R_{(5)} \tag{58}$$

where $\gamma = \det[\gamma_{\hat\mu\hat\nu}]$ and $R_{(5)}$ is the 5-D curvature scalar computed from the metric (57). The field equations for $R_{\mu\nu}$ are then[22]:

$$R^{\mu\nu} - \frac{1}{2} g^{\mu\nu} R = \frac{\kappa^2 \phi_K^2}{2} T^{\mu\nu}_{(em)} - \frac{1}{\phi_K} [\nabla_\mu (\partial_\nu \phi_K) - g_{\mu\nu} \Box_{(4)} \phi_K] \tag{59}$$

where

$$\kappa^2 = \frac{16\pi G}{c^4} \tag{60}$$

In equation (59) the electromagnetic energy-momentum tensor is given by

$$T^{\mu\nu}_{(em)} = \frac{1}{\mu_0}\left(F^{\mu\lambda}F^{\nu}{}_{\lambda} - \frac{1}{4}g^{\mu\nu}F_{\lambda\sigma}F^{\lambda\sigma}\right) \tag{61}$$

In addition, Klein's theory results in (mixed) equations for the EM and scalar fields:

$$\nabla^{\mu}F_{\mu\nu} = -3\frac{1}{\phi_K}\frac{\partial \phi_K}{\partial x^{\mu}}F_{\mu\nu} \tag{62}$$

$$\Box_{(4)}\phi_K = \frac{\kappa^2 \phi_K^3}{4}F_{\mu\nu}F^{\mu\nu} \tag{63}$$

If the scalar field $\phi_K$ is set to one, as in Klein's original formulation, the following equations are obtained:

$$G_{\mu\nu} = \frac{8\pi G}{c^4}T^{(em)}_{\mu\nu} \tag{64}$$

$$\nabla^{\mu}F_{\mu\nu} = 0 \tag{65}$$

However, it has been shown[22] that the condition $\phi_K = 1$ is only consistent with equation (61) when $F_{\mu\nu}F^{\mu\nu} = 0$.

In addition to the 5-D theory described above, Klein introduced a variation which was intended to explain the non-physicality of the fifth dimension. Klein assumed that the fifth dimension was spatial, and assigned it a circular topology and a very small scale. Under the first assumption, any quantity $f(\mathbf{r},t,x^5)$ becomes periodic:

$$f(\mathbf{r},t,x_5) = f(\mathbf{r},t,x_5 + 2\pi L) \tag{66}$$

where $L$ is the radius of the fifth dimension. Hence all the fields may be expanded in Fourier series

$$g_{\mu\nu}(\mathbf{r},t,x_5) = \sum_{-\infty}^{+\infty}g_{\mu\nu}^{(n)}(\mathbf{r})e^{inx_5/L} \tag{67}$$

$$A_\mu(\mathbf{r},t,x_5) = \sum_{-\infty}^{+\infty} A_\mu^{(n)}(\mathbf{r}) e^{inx_5/L} \tag{68}$$

$$\phi_K(\mathbf{r},t,x_5) = \sum_{-\infty}^{+\infty} \phi_K^{(n)}(\mathbf{r}) e^{inx_5/L} \tag{69}$$

where $n$ refers to the $n$-th Fourier component. If $L$ is small enough, of course, it will not be apparent to observers residing in a 4-D universe. The terms $g_{\mu\nu}^{(n)}$, $A_\mu^{(n)}$ and $\phi_K^{(n)}$ satisfy the d'Alembert equations

$$\Box_{(5)} g_{\mu\nu}^{(n)} + \frac{n^2}{L^2} g_{\mu\nu}^{(n)} = 0 \tag{70}$$

$$\Box_{(5)} A_\mu^{(n)} + \frac{n^2}{L^2} A_\mu^{(n)} = 0 \tag{71}$$

$$\Box_{(5)} \phi_K^{(n)} + \frac{n^2}{L^2} \phi_K^{(n)} = 0 \tag{72}$$

where $\Box_{(5)}$ is the 5-D d'Alembertian defined by $\Box_{(5)} = \Box_{(4)} - \partial^2/\partial x_5^2$. Comparison of these potentials with the Klein-Gordon equation has been interpreted as signifying that photon masses are associated with these fields, these being given by

$$m_n \sim \frac{n}{L} \tag{73}$$

In Kaluza's formulation expansions in the form of equations (67) to (69) are not possible, since the cylinder condition forbids the dependence of the terms $g_{\mu\nu}$, $A_\mu$ and $\psi$ on $x_5$.

## IV. KALUZA VS. KLEIN

It is clear from the previous two sections that the Kaluza and Klein 5-D formulations are distinct, and produce very different results in terms of the field equations describing gravitation and electromagnetism. Kaluza's theory results in three sets of equations, the first of which (37) is entirely equivalent to the 4-D Einstein GR equations, the second (38) of which is equivalent to Maxwell's EM equations, and the third (39) which describes a scalar field. The source terms for these three sets of field equations is the 5-D energy-momentum array of equations (40), and the sources are expressed in terms of the material velocities, the velocity of light $c$, and the "fifth" velocity $c_5$. In Klein's theory, one source for the GR field equations in equations (59) and (64) is the "energy-momentum tensor" of the electromagnetic field, the other - in equation (59) - being a term involving the scalar field. This so-called energy-momentum tensor is an entity that is clearly different in content from the matter energy-momentum tensor in Kaluza's theory. What equations (59) and (64) imply is that - in some manner - the sources for the gravitational field has been "geometrized", and are now expressed in terms of the components of the electromagnetic field, $F_{\mu\nu}$, and of the scalar field $\phi_K$, so that in essence the electromagnetic and scalar fields generate the gravitational field. This result is contradicted by available experimental data, which show that the interaction between the electromagnetic and gravitational fields at the present (cosmological) time is extremely weak[23].

The difference between the two theories is primarily a geometrical one, in that Kaluza simply added the fifth dimension $x^5$ without making any changes to the metric structure (i.e. he did not require that the 5-D line element $d\sigma$ be invariant under coordinate transformations, but rather retained the invariance of the original 4-D element $ds$). Although this might seem to be a minor difference, it produces very different metric tensors. In Kaluza's theory, the 5-D metric is simply

$$g_{\hat{\mu}\hat{\nu}} = \begin{bmatrix} g_{\mu\nu} & 2\alpha A_\mu \\ 2\alpha A_\nu & 2\psi \end{bmatrix} \quad (74)$$

in which the $g_{\mu\nu}$ terms are the same as those in the 4-D Einstein GR formulation, the four vectors $A_\mu$ are the electromagnetic potentials, and $\psi$ is the scalar potential. It is apparent from (74) that the gravitational, electromagnetic, and scalar fields are separate entities, and there is no interaction (at least in terms of the fields) between them. On the other hand, in Klein's requirement that $d\sigma$ be invariant results in the metric

$$\gamma_{\hat{\mu}\hat{\nu}} = \begin{bmatrix} g_{\mu\nu} + \kappa^2 \phi_K^2 A_\mu A_\nu & \kappa^2 \phi_K^2 A_\mu \\ \kappa^2 \phi_K^2 A_\nu & \phi_K^2 \end{bmatrix} \quad (75)$$

It is not surprising, therefore, that the resultant field equations differ substantially between the two formulations. In the Klein theory the invariance requirement on the line element $d\sigma$ results in metric component terms in which the three fields (gravitation, EM, and the scalar field) are intermixed. This explains why the terms $F_{\mu\nu}$ appear as if they were sources in the equations for the gravitational field. Additionally, the electromagnetic and scalar fields are intermixed in the field equations (62) and (63).

The spatial structure and invariance properties of the Kaluza and Klein theories are discussed by Einstein and Bergmann[20], Jordan[24], and Ivanov and Prodanov[15]. Although there is no overt consensus as to the which of the two theories more accurately describes 5-D gravity and electromagnetism, in practice it is the Klein theory that has come to predominate, probably because of its apparent mathematical rigor (in comparison with Kaluza's theory, at least). However, there is no theoretical or empirical reason why one should demand the invariance of the 5-D line element $d\sigma$, rather than retain the 4-D invariance of $ds$, as in Kaluza's theory, and the fact that the former demand results in field equations in which the gravitational, electromagnetic and scalar fields are intermixed and interdependent suggests that this constraint is unphysical, and therefore invalid.

## V. SECTORIAL APPROACH TO KALUZA THEORY

The problems in Kaluza's and Klein's versions of the 5-D formulation of gravitation and electromagnetism appear to stem in part from the full application of the cylinder condition (meaning that $g_{\hat{\mu}\hat{\nu},5} = 0$ or $\gamma_{\hat{\mu}\hat{\nu},5} = 0$ applies to all components of the 5-D metric), and in part from the construction of the 5-D source array by simple extension from the corresponding 4-D tensor. In addition, as discussed in the previous section, the problems with Klein's formulation are due to the requirement of the invariance of the 5-D line element $d\sigma$. It is possible to obtain a more consistent 5-D theory by partially reducing the cylinder constraint, and by addressing the transformation properties of different sectors of the metric (70) of Kaluza's theory in a sectorial manner, such that each sector transforms according to a particular prescription, without requiring the 5-D line element $d\sigma$ to be invariant. This is in the spirit of Gonzalez-Mestres' concept of sectorial Lorentz invariance affecting different divisions of spacetime[21].

### A. Relaxed cylinder condition

Under the relaxed cylinder condition, the requirement that $g_{\hat{\mu}\hat{\nu},5} = 0$ will only apply to the four dimensional metric term $g_{\mu\nu}$, while the terms $g_{\mu 5}$ and $g_{55}$ are allowed to depend on $x^5$:

$$g_{\mu\nu} = g_{\mu\nu}(x^1, x^2, x^3, ct)$$

$$g_{\mu 5} = g_{\mu 5}(x^1, x^2, x^3, ct, x^5) \tag{76}$$

$$g_{55} = g_{55}(x^1, x^2, x^3, ct, x^5)$$

It can be seen that only those components of the metric that represent gravitation are affected by the constraint. The coordinate dependence of the field terms is thus

$$R_{\mu\nu} = R_{\mu\nu}(x^1, x^2, x^3, ct)$$

$$A_\mu = A_\mu(x^1, x^2, x^3, ct, x^5) \tag{77}$$

$$\psi = \psi(x^1, x^2, x^3, ct, x^5)$$

## B. Coordinate transformation properties

The next step in the sectorial approach is to specify the transformation properties of the different sectors of the Kaluza 5-D metric (70). Obviously it is desirable to retain the Lorentz covariance of the sector of the 5-D array that corresponds to the 4-D tensor $g_{\mu\nu}$, in order to keep the gravitational field equations (27) intact. Similarly, (4-D vector) Lorentz covariance of the sector(s) that correspond to the 4-vector $[A_x, A_y, A_z, \phi_e/c]^T$ must be retained, so that Maxwell's equations should still apply, at least on the 4-D brane. Insofar as the scalar potential is concerned, since global 5-D covariance with respect to Lorentz transformations is not required, the transformation of the coordinate $x_5$ can be chosen to be simply $x_5' = x_5$, and consequently $\psi$ will transform according to

$$\psi'(x', y', z', t', x_5') = \psi(x, y, z, t, x_5) \tag{78}$$

## C. Field array structure and particle confinement

The structure of the gravitational, electromagnetic and scalar sectors of the field array (36) in relation to the coordinates $x, y, z, ct$ and $x_5$ can be used to interpret the location (in terms of confinement to particular sectors of space, e.g. brane vs. bulk) of the carriers of the three forces. The gravitational sector of the array occupies the first four rows and columns $(x, y, z, ct) \times (x, y, z, ct)$, so it is reasonable to interpret this as meaning that spin-2 gravitons are confined to the 4-D brane. The electromagnetic terms are located in the fifth row and column, corresponding to $x_5 \times (x, y, z, ct)$ space, and this may be interpreted as meaning that photons exist both on the 4-D brane and in the 5-D bulk. This could mean that either (a) there are two classes of photons (brane photons and bulk photons), and that the former propagate at velocity $c$ while the latter propagate at velocity $c_5$, or (b) that photons are in some manner 'hybrid' particles, existing partially on the brane and partially in the bulk. This would require (a) postulating the existence of a new class of particles, or (b) modifying the properties of ordinary photons. Neither of these options is particularly appealing in light of our current understanding of the EM field. In an alternative interpretation of the field array structure, the EM sector might simply occupy the (vector) space $(x, y, z, ct)$, leaving the $x_5$ sector (space) for the scalar field alone. In this scenario, adopted here, both the gravitational and electromagnetic fields are confined to the brane, and only the scalar field occupies the bulk. In is apparent that in this interpretation, the EM field potentials no longer required to depend on the coordinate $x_5$, since the field is restricted (confined) to the 4-D brane. Thus $A_\mu = A_\mu(x^1, x^2, x^3, ct)$ rather than $A_\mu = A_\mu(x^1, x^2, x^3, ct, x^5)$.

## D. Source array structure and coupling to source terms

In the array (40) the source terms for the gravitational field equations ($T_{\mu\nu}$) are just the components of the 4-D energy momentum tensor for dust particles. Kaluza assumed that - by analogy - the source terms for the electromagnetic and scalar fields could be written in the same manner (i.e. in the forms $\rho_m c_5 u_\mu$ and $\rho_m c_5^2$). However, this assumption introduces problems that have been discussed in the literature (Ponce de Leon, 2002; Goenner, 2004). The foremost of these, with respect to the present theory, is the fact that this implementation causes $c_5$ to depend on the charge to mass ratio $\rho_e/\rho_m$ of the source particles. In other words, $c_5$ for electrons would be different from $c_5$ for quarks. This obviously contradicts the proposition made above that $c_5$ is a property of the bulk medium. Some of these problems presumably arise from the fact that coupling between field and source terms in relativistic gravitation is different from that between EM field and source terms, in part because in gravitation the source "charge" (the active gravitational mass) is the same as the inertial mass, whereas in EM the source charge is an entirely different entity. More fundamentally, the problems arise from the assumption that the extension of the 4-D source tensor to 5-D could be accomplished in the manner describe above.

Thus, rather than to try to repair the gravitational and EM coupling problems inherent in Kaluza's source 5-D source array, it is simpler to relate the gravitational and EM field sectors of the field and source arrays separately, including the known 4-D coupling mechanisms for each type of field. In this manner, the coupling of the gravitational field terms in equation (31) to the 4-D matter energy-momentum tensor remains intact. The electromagnetic field terms are now related to the 4-D electric current vector $J_\mu$ through a coupling constant $\gamma_e$ such that the field equations for the EM potentials $A_\mu$ are identical with their Maxwell counterparts. The scalar field is coupled to a scalar "charge" $\rho_s$ whose nature is unknown, through a coupling constant $\gamma_s$. The source array $T_{\hat\mu\hat\nu}$ is now

$$T_{\hat\mu\hat\nu} = \begin{bmatrix} \rho_m u_x^2 & \rho_m u_x u_y & \rho_m u_x u_z & \rho_m c u_x & \gamma_e c_5 J_x \\ \rho_m u_y u_x & \rho_m u_y^2 & \rho_m u_y u_z & \rho_m c u_y & \gamma_e c_5 J_y \\ \rho_m u_z u_x & \rho_m u_z u_y & \rho_m u_z^2 & \rho_m c u_z & \gamma_e c_5 J_z \\ \rho_m c u_x & \rho_m c u_y & \rho_m c u_z & \rho_m c^2 & \gamma_e c c_5 \\ \gamma_e c_5 J_x & \gamma_e c_5 J_y & \gamma_e c_5 J_z & \gamma_e c c_5 & \gamma_s c_5^2 \end{bmatrix} \qquad (79)$$

It is instructive to examine the structure of this array in the case when $u_i \ll c$. The terms that are second order in the source matter velocities are then negligible, and the array becomes

$$T_{\hat{\mu}\hat{\nu}} = \begin{bmatrix} 0 & 0 & 0 & \rho_m c u_x & \gamma_e c_5 J_x \\ 0 & 0 & 0 & \rho_m c u_y & \gamma_e c_5 J_y \\ 0 & 0 & 0 & \rho_m c u_z & \gamma_e c_5 J_z \\ \rho_m c u_x & \rho_m c u_y & \rho c u_z & \rho_m c^2 & \gamma_e c c_5 \\ \gamma_e c_5 J_x & \gamma_e c_5 J_y & \gamma_e c_5 J_z & \gamma_e c c_5 & \gamma_s c_5^2 \end{bmatrix} \quad (80)$$

The (doubly linearized) field potential array corresponding to $T_{\hat{\mu}\hat{\nu}}$ in equation (80) is

$$\bar{h}_{\hat{\mu}\hat{\nu}} = \begin{bmatrix} 0 & 0 & 0 & \bar{h}_{14} & 2\alpha A_x \\ 0 & 0 & 0 & \bar{h}_{24} & 2\alpha A_y \\ 0 & 0 & 0 & \bar{h}_{34} & 2\alpha A_z \\ \bar{h}_{14} & \bar{h}_{24} & \bar{h}_{34} & \bar{h}_{44} & 2\alpha \phi_e/c \\ 2\alpha A_x & 2\alpha A_y & 2\alpha A_z & 2\alpha \phi_e/c & 2\psi \end{bmatrix} \quad (81)$$

The 4-D field tensor equations for the gravitational sector $\bar{h}_{\mu\nu} = 2\kappa T_{\mu\nu}$ now correspond to the vector formulation of gravity (gravitoelectromagnetism)[25], which can be expressed in the equations for the EM potentials

$$\frac{1}{c^2}\frac{\partial^2 \mathbf{A}_g}{\partial t^2} - \nabla^2 \mathbf{A}_g = -\frac{4\pi G}{c^2}\mathbf{J}_m \quad (82)$$

$$\frac{1}{c^2}\frac{\partial^2 \phi_g}{\partial t^2} - \nabla^2 \phi_g = -4\pi G \rho_m \quad (83)$$

in which $\mathbf{A}_g$ the "gravitomagnetic" vector potential, and $\phi_g$ is the "gravitoelectric" scalar potential. The three components of $\mathbf{A}_g$, together with $\phi_g$ correspond to the field terms $\bar{h}_{\mu 4}$ in equation (81). In this formulation, the field equations for gravitation and electromagnetism are entirely homologous, and therefore the corresponding terms in the gravitational and electromagnetic source tensors should also have the same form. It is thus apparent that the source array (80) is still incorrect. The first three entries in the fifth column and row of $T_{\hat{\mu}\hat{\nu}}$ should have the same form as the first three entries in the fourth column and row, i.e. the velocity associated with the electromagnetic source terms should be $c$ rather than $c_5$. The next term, $T_{54}$, should therefore be $\gamma_e c^2$ by analogy with the (gravitational) $T_{44}$ term, and - finally - the $T_{55}$ term should be $\gamma_s c_5^2 \rho_s$. Thus the corrected source array reads

$$T_{\hat{\mu}\hat{\nu}} = \begin{bmatrix} \rho_m u_x^2 & \rho_m u_x u_y & \rho_m u_x u_z & \rho_m c u_x & \gamma_e c J_x \\ \rho_m u_y u_x & \rho_m u_y^2 & \rho_m u_y u_z & \rho_m c u_y & \gamma_e c J_y \\ \rho_m u_z u_x & \rho_m u_z u_y & \rho_m u_z^2 & \rho_m c u_z & \gamma_e c J_z \\ \rho_m c u_x & \rho_m c u_y & \rho_m c u_z & \rho_m c^2 & \gamma_e c^2 \rho_e \\ \gamma_e c J_x & \gamma_e c J_y & \gamma_e c J_z & \gamma_e c^2 \rho_e & \gamma_s c_5^2 \rho_s \end{bmatrix} \quad (84)$$

The sectorial organization of this array is now in full conformity with that of the field potential arrays (36) and (81).

The validity of the analogy between the gravitoelectromagnetic formulation of relativistic gravity and Maxwell's equations could, of course, be questioned if the former theory were only a crude approximation of relativistic gravity. However, it has been demonstrated that the prediction of this theory for the advance of Mercury's perihelion is identical to that obtained via Einstein's formulation of general relativity using the Schwarzschild metric[26]. In addition, the gravitoelectromagnetic formulation has been employed to obtain a general expression for the clock effect, which involves the determination of the difference in the orbital periods of two clocks moving in the opposite directions along a circular equatorial orbit around a central rotating mass[27,28]. The prediction of the gravitoelectromagnetic formulation[29] only differs from the result of the GR calculation (itself obtained in a perturbative manner) by terms of the order of $c^{-4}$.

### E. Field equations

The differential equations that describe the gravitational and electromagnetic fields are written compactly in the form

$$\Box_{(4)} \bar{h}_{\hat{\mu}\hat{\nu}} = 2\kappa T_{\hat{\mu}\hat{\nu}} \quad (85)$$

For the gravitational field, the field potential equations are unchanged:

$$\frac{1}{c^2} \frac{\partial^2 \bar{h}_{\mu\nu}}{\partial t^2} - \nabla^2 \bar{h}_{\mu\nu} = 2\kappa T_{\mu\nu} \quad (86)$$

where the 4-D source tensor is still $T_{\mu\nu} = \rho_m u_\mu u_\nu$. As above, this equation describes the propagation of spin-2 gravitons at the velocity $c$ of light.

The 4-D (brane) EM field potential equations are expressed as

$$\frac{1}{c^2} \frac{\partial^2 A_\mu^{br}}{\partial t^2} - \nabla^2 A_\mu^{br} = -\gamma_e J_\mu^e \quad (87)$$

The coupling constant $\gamma_e$ must be specified as $\gamma_e = -4\pi/c$ in order that the equations for the brane potentials should match their corresponding Maxwell versions. The field equations on the brane are then

$$\frac{1}{c^2}\frac{\partial^2 \phi_e^{br}}{\partial t^2} - \nabla^2 \phi_e^{br} = 4\pi \rho_e \tag{88}$$

$$\frac{1}{c^2}\frac{\partial^2 A_i^{br}}{\partial t^2} - \nabla^2 A_i^{br} = \frac{4\pi}{c} J_i^e \tag{89}$$

The scalar field potential equation requires the 5-D d'Alembertian:

$$\Box_{(5)} = \frac{1}{c_5^2}\frac{\partial^2}{\partial t^2} - \nabla^2 - \frac{\partial^2}{\partial x_5^2} \tag{90}$$

which is valid when $\nabla^2$ is expressed in Cartesian coordinates. The field equation is thus

$$\frac{1}{c_5^2}\frac{\partial^2 \psi}{\partial t^2} - \nabla^2 \psi - \frac{\partial^2 \psi}{\partial x_5^2} = \gamma_s \rho_s \delta(x_5) \tag{91}$$

The source particles for the scalar field reside on the brane, and emit scalar bosons directly into the bulk. These bosons propagate at velocity $c_5$. The values of the velocity $c_5$ and coupling constant $\gamma_s$ are, unfortunately, unknown.

### F. Equation of motion

In the sectorial approach proposed here, each sector can be considered to contribute separately to the equations of motion (or force equations). On the brane, the contributions of gravity and electromagnetism are expressed (as forces) in the form

$$f_\mu^{(g)} = m_{tpg} \Gamma^\mu{}_{\sigma\nu} u^\sigma u^\nu \tag{92}$$

$$f_\mu^{(e)} = q_{te} F_{\mu\sigma} u^\sigma \tag{93}$$

Here $m_{tpg}$ is its passive gravitational "charge" of a test particle (equal to its inertial mass $m_{ti}$) and $q_{te}$ is its EM charge. The first of these two equations is just the gravitational force due to spin-2 gravitons, while the second is the Lorentz force due to spin-1 photons.

Since scalar bosons are confined to the bulk, the scalar field does not exert a force on the brane upon which its source particles are located. However, in a configuration in which there are two parallel branes separated by a distance $d_5$, the scalar field produces a force along the opposing brane, as well as an inter-brane force directed along $x_5$. In other words, scalar bosons emitted by particles on brane 1 propagate through the bulk at velocity $c_5$ and, upon reaching brane 2, are captured by massive particles there. The four-force directed along the second brane, and is given by

$$f_\mu^{s2}(x_\mu, d_5) = -q_{ts2}\left[\frac{\partial \psi(x_\mu, x_5)}{\partial x_\mu}\right]_{x_5=d_5} \quad (93)$$

where $q_{ts2}$ is the scalar charge of a test particle on the second brane. The fifth (inter-brane) component oriented along the $x_5$ axis between the two branes is given by

$$f_5^{s12}(x_\mu, d_5) = -q_{ts2}\left[\frac{\partial \psi(x_\mu, x_5)}{\partial x_5}\right]_{x_5=d_5} \quad (94)$$

It is assumed here that these two components of the scalar force are attractive, and could therefore serve to bind particle-antiparticle pairs across the bulk. Thus quark-antiquark, electron-positron, and even photon-antiphoton pairs could be formed due to the scalar field, provided the photon active and passive scalar charges are non-zero. Since open strings are required to terminate on branes[30,31], particle-antiparticle pairs bound by the scalar force may be considered to be analogous to open strings.

### G. Solution to the scalar field equation for a static point source

For a static point scalar charge $q_s$ source located at the origin on the brane, the field equation for the scalar field reduces to

$$\frac{\partial^2 \psi}{\partial r^2} + \frac{1}{r}\frac{\partial \psi}{\partial r} + \frac{\partial^2 \psi}{\partial x_5^2} = -\kappa_s c_5^2 q_s \delta(r') \quad (95)$$

The quasi-spherical 5-D Laplacian in this expression reflects the "cylindrical" aspect of the 5-D space. In this equation, $r$ is the spherical radius measured from the origin on the brane. The solution to this equation is

$$\psi(r, x_5) = \frac{q_s}{(r^2 + x_5^2)^{1/2}} \quad (96)$$

For a test particle located on the second brane, the (inter-brane) force per unit test particle charge directed along the $x_5$ axis due to the potential in (96) is

$$f_5^{s12}(r,d_5) = -\left[\frac{\partial \psi}{\partial x_5}\right]_{x_5=d_5} = \frac{q_s d_5}{(r^2+d_5^2)^{3/2}} \tag{97}$$

This force is strongest at a point directly opposite the source particle, and can thus serve to bind particles together across the brane. The scalar force directed along the second brane is given by

$$f_r^{s2}(r,d_5) = -\left[\frac{\partial \psi}{\partial r}\right]_{x_5=d_5} = \frac{q_s r}{(r^2+d_5^2)^{3/2}} \tag{98}$$

This force would tend to attract a test particle on the opposing brane to a position directly across from the source particle. However, these two components of the scalar force do not act in such a way as to stabilize or confine particles to the brane. Confinement is presumably mediated by a separate mechanism.

### H. Compactification

The periodic compactification scheme proposed by Klein is not compatible with the solution to the scalar potential equations for the simple source configuration obtained above. This solution requires a linear separation between parallel branes. Klein's original circular compactification regime presumably arose from the need to show that the fifth dimension is in some manner analogous to the three dimensions of ordinary space (i.e. the smallness of $x_5$ is the result of the "curling up" of an ordinary dimension). An alternative scheme was suggested by Ehrenfest and Uhlenbeck[32] shortly after the appearance of Klein's papers[3]. They proposed that a (spacelike) fifth dimension might extend "in a ring or between two walls". The latter of the two options can be understood in the present context as meaning that a spacelike $x_5$ is simply "short" (i.e. non-circular, non-periodic, and orthogonal to $x, y, z$ and $t$), and extends between two 4-D branes (walls). Thus the fifth dimension can be considered to be "compact" rather than "compactified" (permanently unextended rather than shrunken from an extended dimension). This configuration is compatible with recent supergravity and cosmological theories.

### VI. DISCUSSION

In the sectorial formulation presented here there is no Kaluza "miracle", through which Maxwell's equations magically arise out of the linearized Einstein equations through the addition of a fifth spatial dimension. Rather, the appearance of Maxwell's equations in the 5-D formulation is a natural consequence of the fundamental similarity between the mathematical structure of the field equations of linearized gravity and EM. This similarity is manifest when one compares the gravitoelectromagnetic (vector field) formulation of gravity with Maxwell's field equations for electromagnetism. In fact, the field equations of electromagnetism may also be expressed in tensor form, analogous to equations (86) above for linearized gravity, as

$$\frac{1}{c^2}\frac{\partial^2 \bar{e}_{\mu\nu}}{\partial t^2} - \nabla^2 \bar{e}_{\mu\nu} = 2\gamma_{em} Q_{\mu\nu} \tag{99}$$

with the proviso that the field terms $\bar{e}_{ij}$ and $Q_{ij}$ are zero for $i, j = 1, 2, 3$. Here $\gamma_{em}$ is the electromagnetic coupling constant, $\bar{e}_{\mu\nu}$ are the (tensor) EM field terms, and the $Q_{\mu\nu}$ are the sources. The $\bar{e}_{\mu 4}$ terms correspond to the scalar and vector EM potentials $\phi_e$ and $A_i$, and the $Q_{\mu 4}$ terms to the electric charge and current density components $\rho_e$ and $J_i$. The analogy between Maxwell's equations for electromagnetism and the linearized gravitational equations was first noticed by Thirring[33] in 1918. In a recent publication Verbin and Nielsen[34] have suggested that Kaluza's original 5-D formulation may have been inspired by Thirring's paper.

The smaller miracle of Kaluza's theory, the appearance of an additional (scalar) field or force, is retained, and is the principal result of the sectorial theory. It may still be said that the sectorial theory is a "unified" field theory, even though the field equations for gravity and electromagnetism are independent, but the unification is purely mathematical in that the two sets of equations have the same form. The scalar field (91) is also independent of the EM and gravitational fields, and exists in the bulk where the primary fields are absent. Its function could be to bind particles located on the 4-D matter brane with anti-particles located on the 4-D antimatter brane.

A potentially important clarification of Kaluza's theory is the observation that the velocity $c_5$ is not the velocity of charged source particles, but rather a property of the 5-D bulk ("vacuum"), analogous to the velocity of light. As such, it characterizes the propagation of massless scalar bosons in the bulk. Unfortunately, the sectorial formulation presented here does not predict the value of $c_5$, nor that of the interbrane distance $d_5$. It seems likely that these two values are related. The velocity $c_5$ could be either subluminal or superluminal, since the manifestations of the scalar force are not directly perceptible to observers residing on the brane, provided the interbrane distance $d_5$ is small enough.

---